\documentclass[prd,superscriptaddress,altaffilletter,amsmath,amssymb,aps,10pt,twocolumn,floatfix]{revtex4-1}

\usepackage{graphicx} 
\usepackage{dcolumn} 
\usepackage{bm} 
\usepackage{blindtext}
\usepackage{microtype}

\usepackage[breaklinks]{hyperref}
\hypersetup{colorlinks=true,linktoc=all,linkcolor=blue,citecolor=blue,urlcolor=blue}

\usepackage{siunitx} 
\DeclareSIUnit\year{yr} 
\DeclareSIUnit\cts{cts} 

\usepackage{etoolbox} 
\apptocmd{\sloppy}{\hbadness 10000\relax}{}{}

\usepackage{isotope} 

\newcommand{\0}{\ensuremath{0\nu\beta\beta}}
\newcommand{\2}{\ensuremath{2\nu\beta\beta}}

\newcommand{\bb}{\ensuremath{\beta}}

\newcommand{\g}{\ensuremath{\gamma}}
\newcommand{\xe}{\ensuremath{\isotope[136]{Xe}}}

\newcommand{\mchi}{\ensuremath{m_\chi}}
\newcommand{\dnll}{\ensuremath{\Delta\mathrm{NLL}\xspace}}

\begin{document}
    \title{Search for MeV Electron Recoils from Dark Matter in EXO-200}
    \author{S.~Al~Kharusi}\affiliation{Physics Department, McGill University, Montreal, Quebec H3A 2T8, Canada}
\author{G.~Anton}\affiliation{Erlangen Centre for Astroparticle Physics (ECAP), Friedrich-Alexander-University Erlangen-N\"urnberg, Erlangen 91058, Germany}
\author{I.~Badhrees}\altaffiliation{Permanent address: King Abdulaziz City for Science and Technology, Riyadh, Saudi Arabia}\affiliation{Physics Department, Carleton University, Ottawa, Ontario K1S 5B6, Canada}
\author{P.S.~Barbeau}\affiliation{Department of Physics, Duke University, and Triangle Universities Nuclear Laboratory (TUNL), Durham, North Carolina 27708, USA}
\author{D.~Beck}\affiliation{Physics Department, University of Illinois, Urbana-Champaign, Illinois 61801, USA}
\author{V.~Belov}\affiliation{Institute for Theoretical and Experimental Physics named by A.I. Alikhanov of National Research Centre ``Kurchatov Institute'', Moscow 117218, Russia}\altaffiliation[Now a division of National Research Center ``Kurchatov Institute'', Moscow 123182, Russia]{}
\author{T.~Bhatta}\altaffiliation{Present address: Department of Physics and Astronomy, University of Kentucky, Lexington, Kentucky 40506, USA}\affiliation{Department of Physics, University of South Dakota, Vermillion, South Dakota 57069, USA}
\author{M.~Breidenbach}\affiliation{SLAC National Accelerator Laboratory, Menlo Park, California 94025, USA}
\author{T.~Brunner}\affiliation{Physics Department, McGill University, Montreal, Quebec H3A 2T8, Canada}\affiliation{TRIUMF, Vancouver, British Columbia V6T 2A3, Canada}
\author{G.F.~Cao}\affiliation{Institute of High Energy Physics, Beijing 100049, China}
% ...
    \collaboration{EXO-200 Collaboration}
    \noaffiliation
    \date{\today}

    \begin{abstract}
        We present a search for electron-recoil signatures from the charged-current absorption of fermionic dark matter using the EXO-200 detector. We report an average electron recoil background rate of \SI{6.8e-4}{\cts\per\kilo\gram\per\year\per\kilo\electronvolt} above \SI{4}{\mega\electronvolt} and find no statistically significant excess over our background projection. Using a total \xe{} exposure of \SI{234.1}{\kilo\gram\year} we exclude new parameter space for the charged-current absorption cross-section for dark matter masses between $\mchi{} = \SIrange[]{2.6}{11.6}{\mega\electronvolt}$ with a minimum of \SI{6e-51}{\centi\meter\squared} at \SI{8.3}{\mega\electronvolt} at the \SI{90}{\percent} confidence level. \\ {} \\ {} \\ 
    \end{abstract}

    \maketitle

    \section{Introduction}
    \label{sec:intro}
    
        The significant astrophysical and cosmological evidence for the existence of dark matter combined with the lack of understanding of its particle nature is among the most pressing problems in fundamental particle physics~\cite{Schumann:2019eaa}. Dark matter is expected to consist of one or more new, beyond the Standard Model (SM) particles, whose observed interaction with the SM to date is limited to gravitational attraction. Liquid xenon (LXe) time projection chambers (TPCs) are ubiquitous in the field of rare event searches and currently provide the most stringent limits on the scattering cross-section of dark matter in the form of weakly interacting massive particle (WIMPs)~\cite{XENON:2018voc, LUX:2016ggv, PandaX-4T:2021bab}. With the current generation of experiments coming online, they will probe a large fraction of the remaining parameter space for dark matter masses of $\mathcal{O}(\SI{100}{\giga\electronvolt})$~\cite{XENON:2020kmp, LUX-ZEPLIN:2018poe, PandaX:2018wtu}.
        
        However, due to the lack of any confirmed detection to date of well-motivated candidates such as WIMPs or axions~\cite{Graham:2015ouw}, alternative dark matter models have received recent theoretical and experimental interest~\cite{Kolb2018}. A new generation of experiments is aiming to explore lower mass WIMP dark matter between \SI{1}{\mega\electronvolt} and \SI{10}{\giga\electronvolt}~\cite{Essig2020}. The need to detect smaller energy deposits becomes an experimental challenge for interactions of dark matter through elastic scattering with nuclei or electrons. In this case, the maximum energy that can be transferred to the detector is the kinetic energy of the dark matter particle. Other interaction mechanisms of dark matter with the SM have been studied, which could produce higher energy deposits, e.g., in the MeV energy range~\cite{Dror:2019dib, Pospelov:2008qx, PhysRevLett.109.251302}. Detectors searching for neutrinoless double beta decay (\0) are well optimized for this energy range and have demonstrated some of the lowest background rates for any detector technology, along with well-understood background models~\cite{Dolinski:2019nrj, EXO-200:2015nta, EXO-200:2015edf}. This makes such detectors well suited to perform searches for dark matter interactions depositing energy in the MeV range, in addition to their primary focus of searching for \0.
        
        As a specific example of a dark matter model producing such events, the EXO-200 data is analyzed to search for charged-current absorption of fermionic dark matter~\cite{Dror:2019dib} by \xe{} nuclei, for which the energy deposited in the detector depends on the mass of the dark matter particle, which can be much greater than its kinetic energy. This direct detection process results in a unique detector signature with energies of $\mathcal{O}(\SI{}{\mega\electronvolt})$, which may have evaded detection in previous direct detection experiments. In this work, we present a search for excess events in the $1-\SI{8}{\mega\electronvolt}$ energy range in the complete EXO-200 data set. 

    \section{The EXO-200 Experiment}
        \label{sec:exo}
        
        Between 2011 and 2018, the EXO-200 collaboration searched for the \0{} of \xe{} with a total exposure of \SI{234.1}{\kilo\gram\year}, setting a lower limit on the \0{} half-life at $T_{1/2} > \SI{3.5e25}{\year}$ at \SI{90}{\percent} CL~\cite{EXO-200:2019rkq}. After finishing the first run of the experiment in February 2014 (Phase I) the detector underwent an upgrade before the second run started in May 2016 (Phase II) with lower noise front-end electronics and an increase in the electric field from \SI{380}{\volt\per\centi\meter} to \SI{567}{\volt\per\centi\meter}. These upgrades led to an improved energy resolution and lower energy threshold in Phase II. This work also uses the same complete EXO-200 dataset, combining Phase I and II.

        The detector consisted of a cylindrical time projection chamber (TPC) filled with liquid xenon (LXe) enriched to \SI{80.6}{\percent} in \isotope[136]{Xe}, with \SI{19.1}{\percent} \isotope[134]{Xe} and the remaining fraction comprised of various other isotopes. The TPC was split into two identical drift regions sharing a common cathode, which along with the surrounding field rings provided an electric field for drifting electrons. Each TPC half had a radius of \SI{\sim 18}{\centi\meter} and a drift length of \SI{\sim 20}{\centi\meter}. The TPC was enclosed by a radio-pure thin-walled copper vessel, submerged in a heat transfer cryofluid (HFE-7000~\cite{HFE}), which provided \SI{\sim 50}{\centi\meter} of passive shielding, and was maintained inside a vacuum-insulated copper cryostat. Another \SI{\sim 25}{\centi\meter} of lead around the cryostat provided additional shielding against external radiation. The detector was located inside a clean room at the Waste Isolation Pilot Plant (WIPP) in Carlsbad, New Mexico, under an overburden of $1624^{+22}_{-21}$ meters water equivalent~\cite{EXO-200:2015edf}. An active muon veto system consisted of plastic scintillator panels surrounding the clean room on four sides, which allowed prompt identification of \SI{>96}{\percent} (\SI{>94}{\percent}) of the cosmic ray muons passing through the setup in Phase I (II)~\cite{phd_mitchell}. Rejecting events coincident with these muons allows suppression of cosmogenic backgrounds~\cite{EXO-200:2015edf}. A more detailed description of the detector design and performance can be found in~\cite{Auger:2012gs, EXO-200:2021srn}.
        
        In either drift region, the ionization and scintillation quanta produced by an interaction were collected by a crossed-wire plane at each anode, and by an array of large area avalanche photo-diodes (LAAPDs) behind the wire planes, respectively. The total reconstructed energy of an event was determined by combining the light and charge signals defining a total combined energy variable, significantly improving the energy resolution relative to either signal channel alone~\cite{EXO-200:2003bso}. In addition, the combination of light and charge signals provided 3D position reconstruction with an x-y-position resolution of \SI{2.6}{\milli\meter} and a z-position resolution of \SI{0.42}{\milli\meter}~\cite{EXO-200:2013xfn}. The z-direction is defined along the electron drift direction, while the x/y directions lie in the plane parallel to the anode. More details about the analysis and event reconstruction can be found in~\cite{EXO:2017poz, EXO-200:2019rkq}.

    \section{Event Signature and Monte-Carlo}
        \label{sec:mc}
    
        The charged-current absorption of a fermionic dark matter particle $\chi$ by a xenon nucleus will induce a $\bb^-$-decay if the dark matter mass can bridge the gap between the masses of the initial and final nuclei and the mass of the outgoing electron (kinematic threshold). In addition, the daughter nucleus will essentially always be produced in an excited state for the isotopes of interest here due to angular momentum selection rules. The event rate for this process is given by~\cite{Dror:2019dib, Dror2020}
        \begin{align}
            R  = & \frac{\rho_\chi}{2m_\chi} \sum_{j,k} N_{T,j} \cdot n_j \cdot \mathcal{F}_k(Z+1,E_{e,k}) \nonumber \\
           & \times \frac{|\vec{p}_{e,k}|}{16\pi m_\chi M^2_{A_j,Z_j}}  \overline{|\mathcal{M}_k|^2}
        \end{align}
        summing over all possible excited states $k$, where \mchi{} is the dark matter mass, $\rho_\chi=\SI{0.3}{\giga\electronvolt\per\centi\meter\cubed}$ is the local dark matter density, $N_{T,j}$ is the number of targets of a given isotope $j$, $n_j$ is the number of neutrons per target, $\mathcal{F}_k$ is the Fermi function, $\vec{p}_{e,k}$ is the momentum of the outgoing \bb{}, $E_{e,k}$ the energy of the emitted \bb{}, $M^2_{A_j, Z_j}$ the mass of nucleus \isotope[A][Z]{X}, and $\overline{|\mathcal{M}_{k}|^2}$ is the spin averaged nucleon level matrix element, all of which are defined in \cite{Dror:2019dib}. The details of the derivation of the rate, the Fermi function, and an expression for the matrix element can be found in~\cite{Dror:2019dib, Dror2020}. Because of the higher kinematic thresholds~\cite{Sergeenkov1981} and smaller isotopic fraction, absorption of fermionic dark matter by \isotope[134]{Xe} is negligible relative to absorption on \isotope[136]{Xe}, in EXO-200. The main channel considered is, therefore, $\chi + \isotope[136][54]{Xe} \rightarrow \isotope[136][55]{Cs}^* + e^-$, in which the cesium nucleus is produced in an $J^P=1^+$ excited state, where $J$ is the total angular momentum and $P$ is the parity of the state. In the absence of a detection, such data can constrain the interaction cross-section of this charged-current absorption process, or equivalently, the effective energy scale, $\Lambda$, for a beyond the SM operator that could mediate the dark matter interaction 
        \begin{equation}
            \frac{1}{\Lambda^2} \equiv \frac{g_R^2}{4 M_{W_R}^2}
        \end{equation}
        where $g_R$ is the $SU(2)_R$ coupling constant of the theory, $M_{W_R} = \frac{1}{2} g_R u$ is the mass of the mediator (similar to the W-boson as the mediator of charged-current interactions via the weak nuclear force), with $u$ being the vacuum expectation value of the scalar field $\phi$ through which the dark matter particle $\chi$ interacts with the SM via a Yukawa type coupling. 
    
        \begin{figure}
            \centering
            \includegraphics[width=\columnwidth]{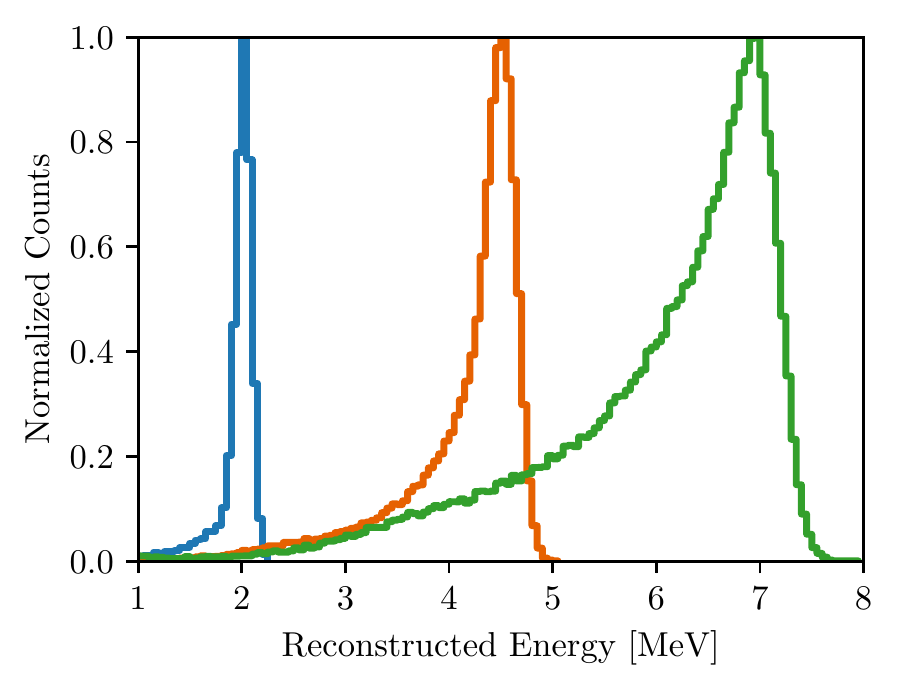}
            \caption{Monte Carlo simulation of the EXO-200 detector response to electrons at energies of \SI{2}{\mega\electronvolt}, \SI{4.5}{\mega\electronvolt} and \SI{7}{\mega\electronvolt} in Phase II. The simulated distribution of the maximum reconstructed ionization energy in an event is shown, where the non-Gaussian shape comes from the increased emission of bremsstrahlung for the higher energy \bb{}s.}
            \label{fig:electrons}
        \end{figure}
    
        A major challenge in modeling these signals in EXO-200 comes from the fact that little data exists on the nuclear level structure of \isotope[136]{Cs}, with no sufficiently fast decaying states currently measured between the lowest lying $1^+$ excited states at \SI{590}{\kilo\electronvolt} and its $5^+$ ground state~\cite{Mccutchan:2018jsr}. Assuming the dark matter absorption preferentially populates the lowest-lying $1^+$ state at \SI{590}{\kilo\electronvolt}, there are only four known states to which the nucleus can decay: three excited states ($J^P = 9^-, 8^-,4^+$), and the ground state ($J^P=5^+$)~\cite{Mccutchan:2018jsr}. In all cases, the decays are highly forbidden and order-of-magnitude Weisskopf estimates for the \SI{590}{\kilo\electronvolt} state give a predicted half-life of $\gg \SI{1}{\second}$. 
        In this case, the relaxation would appear as a secondary, uncorrelated event in the EXO-200 data, with similar signatures for excitations to higher energy nuclear levels of \isotope[136]{Cs}. However, if unknown intermediate levels of appropriate angular momentum and parity exist, then this half-life could be much shorter, allowing a dark matter absorption event to have a characteristic signature including both the primary \bb{} as well as one or more coincident \g{}s. Since neither the decay time nor energy of these \g{}s is currently known, we perform the analysis assuming that the only event signature is the outgoing \bb{}. However, as described in Sec.~\ref{sec:results}, we also consider the case that one or more coincident \g{}s of unknown energy are emitted, and present results that are insensitive to the presence of such \g{}s. 
        We note that the level structure of \isotope[136]{Cs} is under active investigation, with data suggesting the existence of new intermediate levels that may be relevant for such dark matter interactions~\cite{Rebeiro:2016pvi}. In addition, recent shell model calculations predict a decay scheme in which the lowest-lying $1^+$ state relaxes through levels that include an isomeric state with a lifetime of $\mathcal{O}(\SI{100}{\micro\second})$~\cite{Haselschwardt:2020ffr}, which could enable a time-coincidence analysis to separate charged-current absorption events from backgrounds. However, since the existence of these intermediate states has not been conclusively confirmed, we proceed as described above and leave their inclusion in dark matter analyses to future work.

        The detector response to ionizing radiation is modeled by a detailed Monte Carlo (MC) simulation based on Geant4~\cite{GEANT4:2002zbu}. The details of the MC simulation and reconstruction can be found in~\cite{EXO-200:2013xfn, EXO-200:2019rkq}. The absorption event signature of a fermionic dark matter particle is modeled via a single \bb{} with a kinetic energy equal to the dark matter mass $m_\chi$ minus the kinematic threshold energy~\cite{Dror:2019dib}. The transferred kinetic energy of the dark matter particle results in a $\mathcal{O}(\SI{}{\kilo\electronvolt})$ nuclear recoil of the \xe{} atom which falls below the energy threshold and is therefore ignored. For a given dark matter mass the injected energy into the detector will be mono-energetic and produce a peak-like signature in the reconstructed energy spectrum. The maximum reconstructed ionization energy spectrum of \bb{}s at various energies in Phase II is shown in Figure \ref{fig:electrons}. The more energetic the emitted \bb{}, the more bremsstrahlung it emits whose reconstructed energy might fall below the detector threshold, leading to a non-Gaussian low energy tail.

    \section{Analysis} 
        \label{sec:analysis}
    
        The energy region considered here can be divided into two regions based on the relevant backgrounds. In the low energy portion below \SI{3}{\mega\electronvolt}, the dominant background sources are \2{} events and \g{}s from the \isotope[238]{U} and \isotope[232]{Th} decay chains. Above \SI{3}{\mega\electronvolt} the main backgrounds arise from de-excitation \g{}s from cosmogenically produced isotopes. 
        To separate the \bb{} arising from dark matter interactions from backgrounds, one can exploit the fact that single MeV-scale \bb{}s typically deposit most of their kinetic energy at a single location (i.e., in the most energetic reconstructed charge cluster), in contrast to \g-backgrounds that predominantly Compton-scatter and therefore, on average, have less of their reconstructed energy contained in the most energetic cluster. An additional energy variable, defined as the energy of the most energetic charge deposit in an event, captures this topology difference. Therefore, in combination with the total reconstructed event energy, the energy of the maximum reconstructed charge cluster is used as a second dimension in the likelihood fits discussed in Sec.~\ref{sec:results}.

        The fiducial volume and data quality cuts and the corresponding fiducial mass are the same as in~\cite{EXO-200:2019rkq}, resulting in a total exposure of \SI{234.1}{\kilo\gram\year} divided into \SI{117.4}{\kilo\gram\year} and \SI{116.7}{\kilo\gram\year} in Phase I and Phase II, respectively.
    
        Compared to previous \0 analyses from EXO-200~\cite{Auger:2012gs,EXO-200:2019rkq} this analysis 
        \begin{itemize}
            \item extends the energy range up to \SI{8}{\mega\electronvolt}
            \item combines all events into a single dataset irrespective of the number of interaction sites
            \item adds a second analysis variable capturing the energy of the most energetic charge deposit in an event.
        \end{itemize}
    
\begin{figure}
            \centering
            \includegraphics[width=\columnwidth]{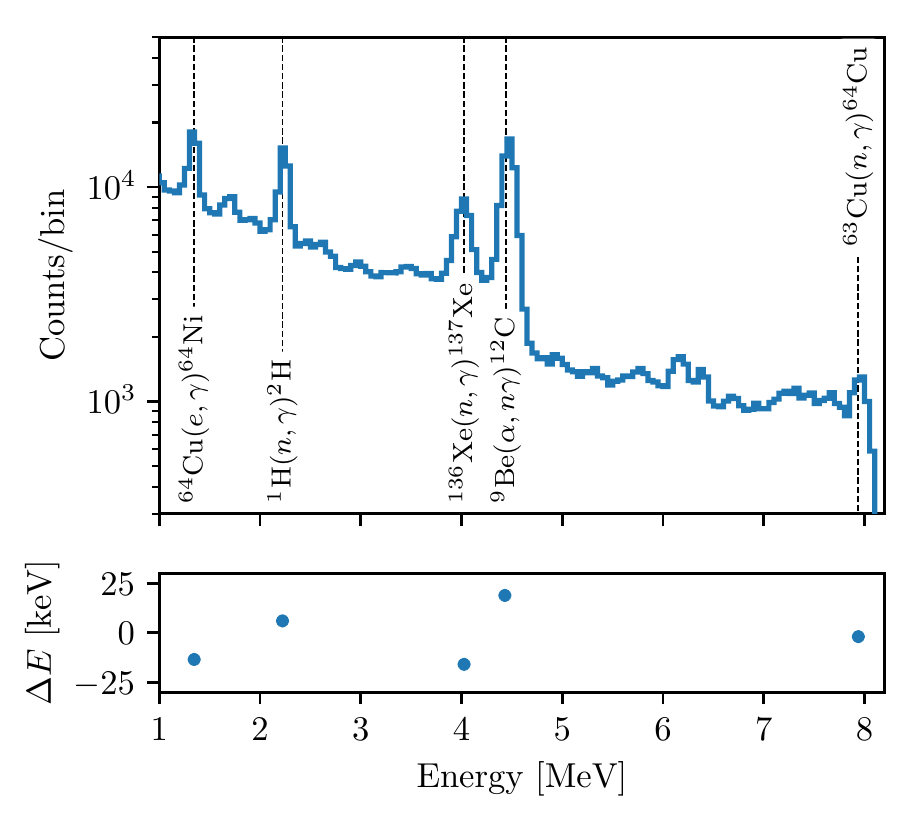}
            \caption{(top) Reconstructed AmBe calibration source energy spectrum in Phase II, demonstrating linear energy response of the EXO-200 detector throughout the energy range of interest. (bottom) Residuals between the measured and known energies $\Delta E$. Calibration errors at the photo-peaks are less than \SI{20}{\kilo\electronvolt}, resulting in an systematic error on the energy scale of \SI{<0.5}{\percent}.}
            \label{fig:energy_calibration}
        \end{figure}

        The energy response of the detector is typically calibrated with external \g{}-sources placed on the outside of the detector near the cathode, using the full absorption peaks resulting from the decays of \isotope[60]{Co} and daughters of \isotope[226]{Ra} and \isotope[228]{Th}. At the end of the EXO-200 livetime, a composite americium beryllium (AmBe) neutron source was used to produce radiogenically activated \isotope[137]{Xe}, whose \bb{}-decay spectral shape was previously studied~\cite{EXO-200:2020wmu}. In addition, the AmBe source calibration provided a high statistics dataset of various other neutron activated by-products whose de-excitation gammas can be used for an energy calibration up to \SI{\sim 8}{\mega\electronvolt} in Phase II only. In addition to the primary \g{} emitted at \SI{4430}{\kilo\electronvolt} ($\isotope[9]{Be} (\alpha,n\g) \isotope[12]{C}$), this dataset includes full absorption peaks at \SI{1346}{\kilo\electronvolt} ($\isotope[64]{Cu} (e,\g) \isotope[64]{Ni}$), \SI{2223}{\kilo\electronvolt} ($\isotope[1]{H} (n,\g) \isotope[2]{H}$), \SI{4025}{\kilo\electronvolt} ($\isotope[136]{Xe} (n,\g) \isotope[137]{Xe}$), and \SI{7916}{\kilo\electronvolt} ($\isotope[63]{Cu} (n,\g) \isotope[64]{Cu}$). The reconstructed energy spectrum resulting from the AmBe source and the corresponding energy calibration results are shown in Figure \ref{fig:energy_calibration}. The lower bound of the energy range considered for this analysis follows previous EXO-200 \0 analyses~\cite{EXO-200:2019rkq}, whereas the upper bound is limited by the highest available energy calibration peak at \SI{7916}{\kilo\electronvolt}.
    
        While the analysis is limited to below \SI{8}{\mega\electronvolt} to ensure the energy calibration is well-understood, the EXO-200 detector is sensitive to higher energy deposits. The low background data contains 18 (41) events above \SI{8}{\mega\electronvolt} in Phase I (II) passing all event selection cuts, but which are not accounted for by any of the components in the EXO-200 background model~\cite{EXO-200:2015edf, EXO-200:2015nta}. No events above \SI{8}{\mega\electronvolt} are present in the veto-tagged data containing only events that were collected between \SI{10}{\micro\second} and \SI{5}{\milli\second} after any of the 29 muon veto panel triggers. This suggests that these events in the low-background data could arise, e.g., from long-lived cosmogenic activation products, or unaccounted for radiogenic processes not present in the veto-coincident data. The possible effect of such a small, unknown background on the results of this analysis will be discussed in Sec.~\ref{sec:systematics}. 
    
        To maximize the signal efficiency for \bb{} events in this energy range, this analysis does not require all ionization signals in an event to be fully reconstructed. Relative to the selection cuts used in \0 searches, this recovers events with multiple bremsstrahlung photons, for which there is a substantial probability that at least one low-energy cluster is detected for which the $x$ and $y$ position can't be fully reconstructed. With all selection criteria included, the Monte Carlo simulation is used to estimate a signal reconstruction efficiency of $\epsilon > \SI{99.7}{\percent}$ for both data-taking phases. Possible systematic errors in this efficiency will be discussed in Sec.~\ref{sec:systematics}.
    
        MC events from all background and signal components, passing all quality cuts, are used to construct individual two-dimensional probability density functions (PDFs) that depend on the reconstructed total energy and the maximum charge cluster energy of an event. The components entering the background model are the same as in~\cite{EXO-200:2019rkq}, whereas the \0{} signal is replaced by a monoenergetic \bb{} PDF (one for every \SI{50}{\kilo\electronvolt} step within the analysis energy window), which is shown in Figure~\ref{fig:electrons}. The search was performed by minimizing a binned negative log likelihood (NLL) function when fitting the combined signal and background model PDFs to the data. Systematic errors are added to the fit as nuisance parameters that are constrained by a normal distribution, with a width corresponding to the size of the systematic error estimated in stand-alone studies (see Sec.~\ref{sec:systematics}). 
        
        Toy datasets are generated from a background-only fit to the low-background data. These toy datasets are fit with the full background plus signal model, and the upper limit on the number of signal counts for each signal PDF at the \SI{90}{\percent} CL is determined from a profile likelihood. The sensitivity is obtained by calculating the median of an ensemble of 1000 upper limits.

    \section{Systematics}
        \label{sec:systematics}
        
        The assessment of systematic errors follows the same general approach as in~\cite{EXO-200:2019rkq} and includes an 
        \begin{enumerate}
            \item uncertainty in the signal-specific detection efficiency caused by discrepancies in the shape of data and MC PDFs
            \item uncertainty in the activity of radon in the LXe as determined in stand-alone studies via measurement of time-correlated decays
            \item uncertainty in the overall event detection efficiency due to possible errors in the estimated event reconstruction and selection efficiencies
        \end{enumerate}
        \begin{figure}
            \centering
            \includegraphics[width=\columnwidth]{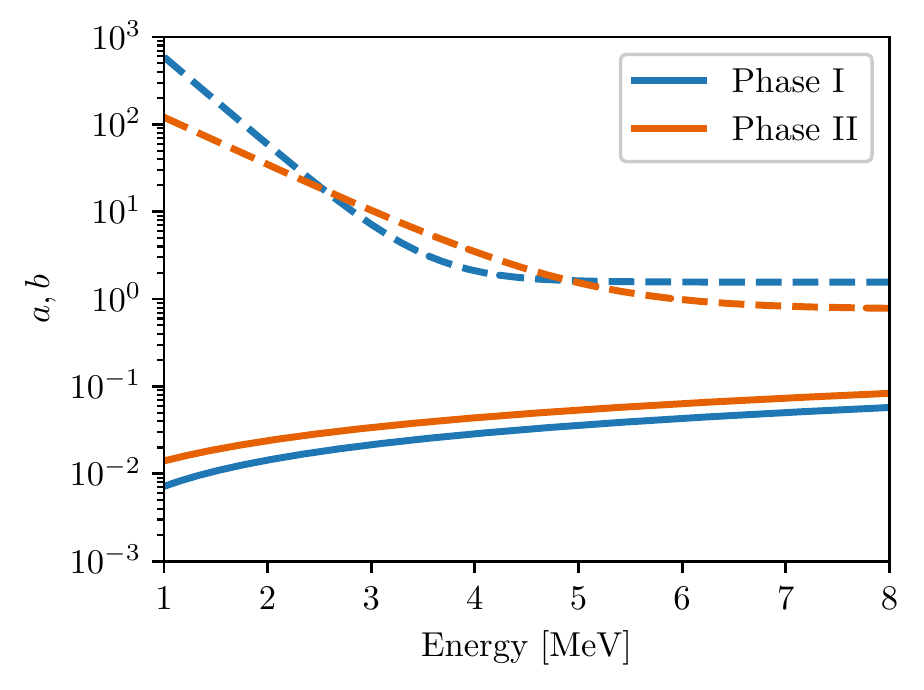}
            \caption{Energy dependence of the parameters $a$ (solid) and $b$ (dashed) in equation \ref{eq:shape_error} for quantifying the systematic error from spectral shape disagreements, which are parameterized by $a(E)=a_1\cdot E + a_2$ and $b(E)=b_1\cdot e^{- b_2\cdot E} + b_3$, respectively.}
            \label{fig:shape_error}
        \end{figure}
    
        The first systematic error arises from shape disagreements between data and MC PDFs and is propagated to the fit as a Gaussian constraint on the normalization of the signal PDF, as was done in~\cite{EXO-200:2019rkq}. We performed a fit to the veto-tagged datasets in Phase I and II and treated the bin-by-bin ratio between the data and the best-fit model in each fit dimension as a possible error, reweighting the shape when building the PDFs for the likelihood fit to the low-background data to correct for the spectral distortions. Toy datasets were generated from these modified background model PDFs with the signal PDF present and were fit against the original unweighted PDFs. The bias in the number of fitted signal counts as a function of the injected number of signal counts is used to construct a conditional Gaussian constraint for the signal-only PDF normalization that has the following functional form 
        \begin{equation}
            \sigma/N = \sqrt{(a\cdot N)^2+ b^2 }/N \label{eq:shape_error}
        \end{equation}
        It consists of a component proportional to the number of signal counts $N$ with a proportionality factor $a$ and a signal-independent component $b$. The signal-independent component quantifies shape errors in the background model that lead to a fixed bias in the number of signal counts, while the signal-dependent term quantifies a relative error that would be expected to scale with the signal size. The energy dependence of the parameters $a$ and $b$ were found to be approximately described by linear and exponential functions, respectively, and is shown in Figure~\ref{fig:shape_error}. For Phase II an average between the bias from the spectral shape error in the muon veto-coincident data and the AmBe calibration data was used. The second uncertainty, quantifying possible error on the Rn activity in the LXe, was estimated in~\cite{EXO-200:2015edf} to be \SI{10}{\percent} and is only applied to the Rn-related backgrounds. For the last item, this analysis assumes the same systematic error for the overall signal efficiency of \SI{3.0}{\percent} (\SI{2.9}{\percent}) for Phase I (II) as was used in ~\cite{EXO-200:2019rkq}. This systematic error may provide a slightly conservative estimate for the reconstruction efficiencies since \bb{}-like events above \SI{3}{\mega\electronvolt} are easier to identify with smaller reconstruction errors due to their higher signal-to-noise.
        
        As mentioned, the observed events above \SI{8}{\mega\electronvolt} suggest that a small background component not included in the existing EXO-200 background model could be present. If such backgrounds arise from e.g., Compton scattering of high energy $\gamma$s, this background component may also extend into the energy region below \SI{8}{\mega\electronvolt} that is considered in this analysis. We estimated the effect that such an unknown background might have on our results by including an additional freely floating flat background PDF into the fit model and reevaluating the sensitivity. The relative difference in the sensitivity with and without this additional background is taken as a systematic error due to the possible incompleteness of the background model and was calculated to be \SI{\sim 0.1}{\percent}. While a small energy dependence is observed, the largest systematic error over all energies is used as a conservative estimate.
    
        The signature of charged-current absorption events is similar to a solar neutrino interaction with the detector, where in the case of the charged-current interaction ($\nu + \isotope[136][]{Xe} \rightarrow \isotope[136][]{Cs}^* + e^-$) \isotope[136]{Cs} might also be produced in an excited state. In contrast, an elastic scattering process will only result in a single \bb{} in the final state. The solar neutrino background from both processes is estimated to produce $\ll1$ event throughout the livetime of the experiment within the entire energy region for this analysis~\cite{Haselschwardt:2020ffr} and can therefore be safely ignored.
        
        Lastly, the systematic error on the energy scale is evaluated by taking the maximum spread in the calibrated energy for various possible calibration functions that interpolate between the measured photo-peaks of known energy (e.g. interpolating functions consisting of 1st or 2nd order polynomials, with or without a constant offset). The uncertainty in our energy calibration was estimated to be \SI{\sim 2}{\percent} (\SI{\sim 0.5}{\percent}) in Phase I (II). The larger error in Phase I arises from the lack of AmBe calibration data, which provides known energy peaks constraining the calibration functions to the \SI{8}{\mega\electronvolt} upper energy threshold. This systematic is propagated as a Gaussian constraint on the \g-scale, which is a freely floating parameter in the likelihood fit that scales the energy of all PDFs representing interactions of \g{} events via a common multiplicative factor. This floating \g-scale can compensate for possible differences between calibrations and low-background data, but the best fit value was found to be consistent with unity within less than \SI{0.2}{\percent}. An additional \bb-scale is introduced into the fit that allows the energy scale of PDFs arising from \bb{}-like events to float independently from \g-like PDFs, which allows the fit to correct possible differences in the energy response between \bb{}s and \g{}s in the detector. However, this parameter was similarly found to be consistent with unity to within less than \SI{1}{\percent}, in good agreement with previous studies~\cite{EXO-200:2019bbx}. Table~\ref{tab:systematics} shows a summary of the systematic errors considered for this analysis.
        \begin{table}
            \centering
            \begin{tabular}{>{\raggedright\arraybackslash}p{3cm}>{\centering\arraybackslash}p{2.5cm}>{\centering\arraybackslash}p{2.5cm}}
                \hline \hline 
                Constraint \hspace{125pt} & Phase I & Phase II \\
                \hline
                Shape error & $\SI{36}{\percent}^*$ & $\SI{54}{\percent}^*$ \\
                \isotope[222]{Rn} & 10\% & 10\% \\ 
                Normalization & 3.0\% & 2.9\% \\ 
                Background model & 0.1\% & 0.1\% \\ 
                Energy scale & 2\% & $<0.5$\% \\ 
                \hline \hline
            \end{tabular}
            \caption{Summary of estimated systematic errors for this analysis in Phase I and II data sets. $^*$This value is evaluated at \SI{4}{\mega\electronvolt}. The systematic error from spectral shape disagreement is energy-dependent and is shown over the entire energy range in Fig.~\ref{fig:shape_error}.}
            \label{tab:systematics}
        \end{table}

    \section{Results}
        \label{sec:results}

        \begin{figure}
            \centering
            \includegraphics[width=\columnwidth]{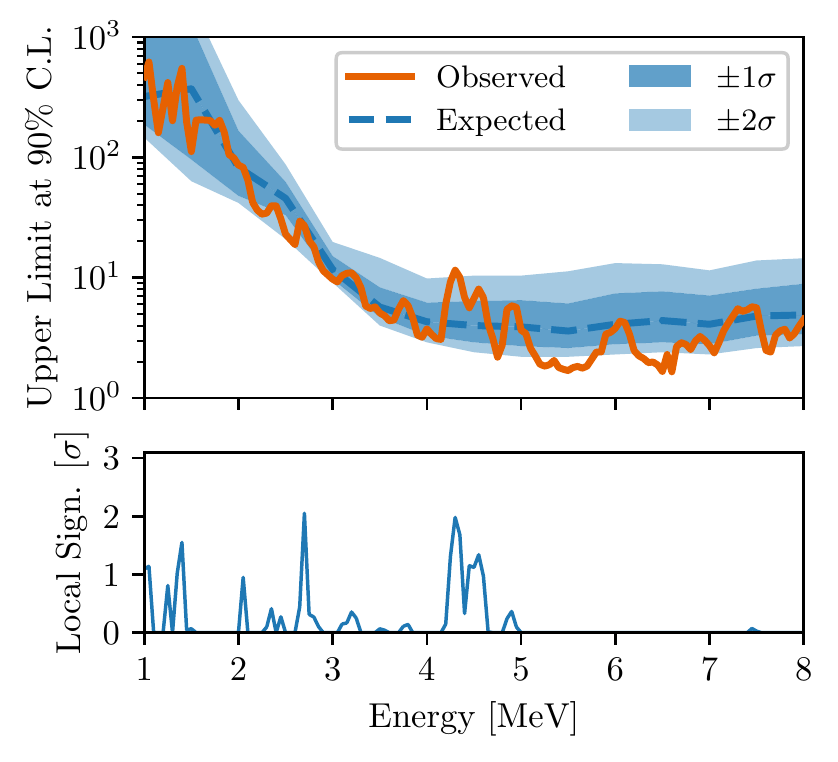}
            \caption{The top panel of the plot shows the upper limit (solid orange) on the number of dark matter events as a function of energy at \SI{90}{\percent} CL. The median sensitivity (dashed blue) and the \SI{\pm 1}{\sigma} (dark blue) and \SI{\pm 2}{\sigma} (light blue) percentiles around the median are also shown. The bottom panel shows the local significance of our observation against the no-signal hypothesis.} 
            \label{fig:limit_combined}
        \end{figure}
        \begin{figure}
            \centering
            \includegraphics[width=\columnwidth]{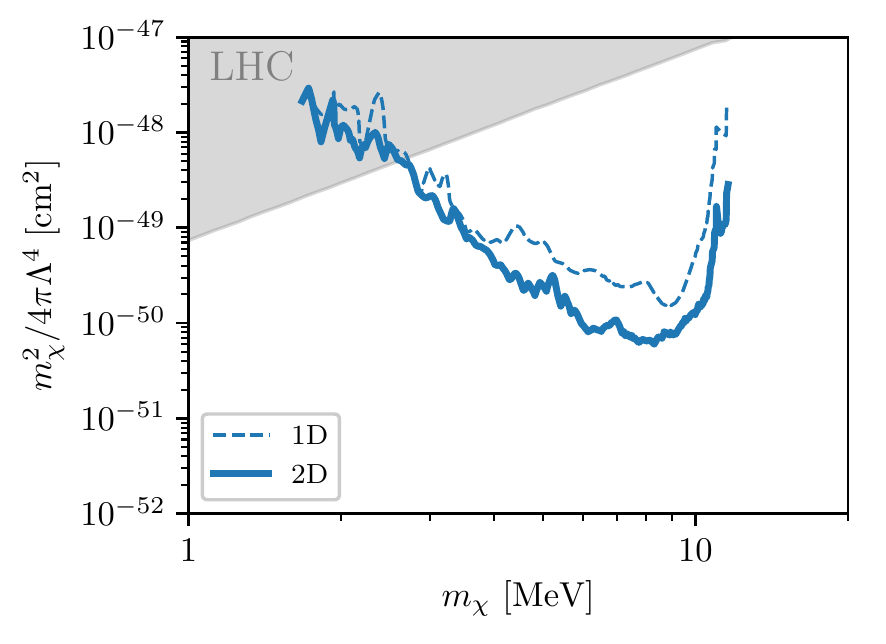}
            \caption{Limit on the absorption cross-section of fermionic dark matter by \xe{} nuclei at the \SI{90}{\percent} CL. We show our main 2D fit result in addition to a de-excitation \g{} agnostic 1D fit result (see text for more details). The grey region is excluded by direct constraint from searches at the LHC~\cite{Belyaev:2018pqr}.}
            \label{fig:fermionic_dm_sens}
        \end{figure}

        Following~\cite{EXO-200:2019rkq}, the likelihood fits are initially performed independently in each data-taking phase, minimizing a binned NLL function and then profiling over the signal PDF at a given energy. Hence, the combined NLL at a given energy is simply the sum of both NLLs, and the efficiency and livetime corrected profile likelihood curves, which are a function of the dark matter interaction cross-section, can be added to obtain the combined upper limit at \SI{90}{\percent} CL. The determination of the \SI{90}{\percent} percentile of the test statistic is assuming that Wilks' approximation holds~\cite{Wilks1938}. A comparison with the obtained upper limit using MC-based test statistic distribution at an example dark matter mass of \SI{5}{\mega\electronvolt} shows that the results of this work produce a \SI{\sim 10}{\percent} weaker limit, and thus the assumption of Wilks' approximation does not have a significant effect on the limit (and may be slightly conservative).
    
        The combined median \SI{90}{\percent} CL sensitivity is similarly determined by combining the profile likelihoods of 1000 toy datasets from each phase. The combined \SI{90}{\percent} upper limit and sensitivity are shown in Fig.~\ref{fig:limit_combined}, together with the \SI{68}{\percent} and \SI{95}{\percent} percentiles around the median sensitivity. We calculate the local $p$-value as the fraction of toy datasets under the no-signal hypothesis whose \dnll{} is at least as large as the one obtained from the data fit. The largest local significance was found to be $\sim 2\sigma$ at \SI{2.7}{\mega\electronvolt}. 
        We, therefore, find no statistically significant evidence for dark matter in our data. Overall, the observed limit lies within the $\pm 2 \sigma$ distribution around the median sensitivity over most of the mass range. In the region between \SI{5}{\mega\electronvolt} and \SI{7}{\mega\electronvolt}, the limit is slightly stronger than the expected $\pm 2 \sigma$ region. This could arise from either a statistical under-fluctuation in the backgrounds at these energies or, possibly, a small overestimate of the background in this region relative to that assumed in the toy Monte Carlo studies.
        
        Figure \ref{fig:fermionic_dm_sens} shows the \SI{90}{\percent} CL exclusion limit on the interaction cross-section as a function dark matter mass \mchi{}. In addition to the primary result using a two-dimensional likelihood fit consisting of the total combined energy and the maximum charge energy in an event, we also performed the fits using only the latter variable. These 1D fits have reduced sensitivity to a signal but provide an analysis that does not strongly depend on the presence of additional de-excitation \g{}s, since the primary \bb{} cluster will contain the largest energy for nearly all events over the mass range considered, resulting in only marginal differences in the signal PDF shape. We also include limits from constraints on these models from collider experiments at the Large Hadron Collider (LHC) at CERN~\cite{Belyaev:2018pqr}, demonstrating that relative to collider searches low-background detectors can perform competitive searches for certain dark matter models. Additional constraints arise from indirect detection searches due to the decay of such dark matter particles that become more stringent at higher dark matter masses~\cite{Essig:2013goa}.

        \begin{figure*}
            \centering
            \includegraphics[width=\textwidth]{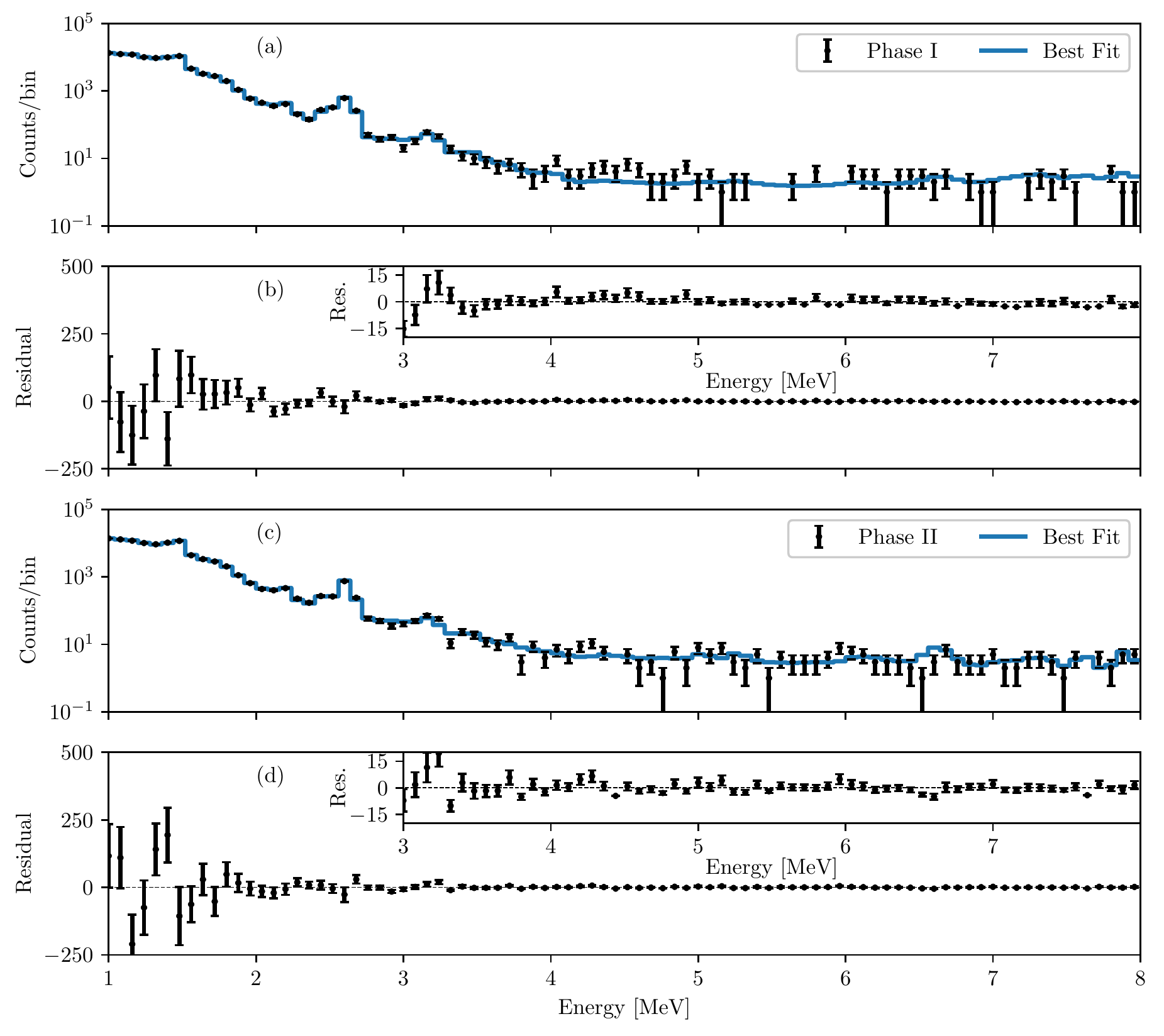}
            \caption{Total combined energy spectrum of the low-background data and the best fit (blue) to the data with a background-only model in Phase I (a) and Phase II (c). The resulting residual number of counts between data and the best from Phase I and Phase II is shown in panels (b) and (d), respectively. No significant disagreement is seen between the data and the background-only model. The average electron-recoil background rate above \SI{4}{\mega\electronvolt} is \SI{4.0e-4}{\cts\per\kilo\gram\per\year\per\kilo\electronvolt} in Phase I and \SI{6.8e-4}{\cts\per\kilo\gram\per\year\per\kilo\electronvolt} in Phase II with a total exposure of \SI{117.4}{\kilo\gram\year} and \SI{116.7}{\kilo\gram\year}, respectively.}
            \label{fig:fit}
        \end{figure*}

        In Fig.~\ref{fig:fit} we provide the total combined energy spectrum of the low-background data from \SI{1}{\mega\electronvolt} to \SI{8}{\mega\electronvolt}, along with the best fit to the data with a background-only model, and the bin-wise residual counts for Phase I and Phase II. The excellent spectral shape agreement and linear detector response to both \bb-like and \g-like events are a result of more than ten years of work to build, run, and understand the EXO-200 detector and its response. These spectra, published here for the first time, may also constrain other physics beyond the SM that is outside the scope of the present work.

        This analysis represents the first search for the absorption of MeV fermionic dark matter in a liquid xenon detector via a charged-current interaction, with a total \xe{} exposure of \SI{234.1}{\kilo\gram\year}. As no statistically significant evidence was observed, we exclude new parameter space for these models at the \SI{90}{\percent} CL for dark matter masses between \SI{2.6}{\mega\electronvolt} and \SI{11.6}{\mega\electronvolt}. These results are complementary to searches for neutral-current absorption of fermionic dark matter reported in~\cite{PandaX:2022osq, Arnquist:2022uxq} and charged-current absorption of light fermionic dark matter in~\cite{PandaX:2022xon}. Furthermore, a detailed understanding of the detector response is demonstrated. The low backgrounds achievable with the liquid xenon technology can enable future rare event searches such as \0, direct detection of WIMP dark matter, or other novel dark matter interaction mechanisms~\cite{nEXO:2021ujk, Avasthi:2021lgy, DARWIN:2016hyl}.

    \section*{Acknowledgements}
        EXO-200 is supported by the DOE and NSF in the U.S., the NSERC in Canada, the SNF in Switzerland, the IBS in Korea, the DFG in Germany, and CAS and ISTCP in China. The EXO-200 data analysis and simulation use resources of the National Energy Research Scientific Computing Center (NERSC). We gratefully acknowledge the KARMEN collaboration for supplying the cosmic-ray veto detectors, as well as the WIPP for their hospitality.
        
    \bibliography{bib}

\end{document}